\documentclass[twocolumn,showpacs,preprintnumbers,amsmath,amssymb]{revtex4}
\usepackage{graphicx}
\usepackage{dcolumn}
\usepackage{bm}

\begin{document}

\title{A model of the T-dependent pseudogap and its competition with superconductivity in copper oxides.}

\author{T. Jarlborg}

\affiliation{
DPMC, University of Geneva, 24 Quai Ernest-Ansermet, CH-1211 Geneva 4,
Switzerland
}

%\date{\today}

\begin{abstract}

Results for pseudogaps are
obtained from a band model, where the stability of the gap depends on the
amplitudes of vibrational displacements, or magnetic moments, and their coupling to electrons.  
A one-particle gap is favored
by normal thermal excitations of phonons or spin waves. Another gap can be generated by spontaneous waves
at lower temperature, if 
the electronic energy gain 
overcomes the elastic/magnetic energy needed for increased amplitudes of
the oscillations. This state is characterized by 
charge or spin density waves. The pseudogap has many features 
in common with the
superconducting gap, and the model lends support 
to the interpretation that the pseudogap is
a precursor of, and competes with, superconducting pairing.

\end{abstract}


\maketitle

%\section{Introduction.}

%\section{Methods.}

The understanding of high-$T_C$ cuprates has increased a lot since the discovery
of these systems more than 20 years ago \cite{mull}, although a convincing mechanism
behind the high superconducting $T_C$ is not found. A pseudogap with 
transition temperature $T^* > T_C$ at low hole doping seems to be characteristic for all
cuprates \cite{tall,timu,dama}.
Ab-initio band calculations for long supercells of copper oxides containing
phonon distortions or spin waves show a partial gap (pseudogap) in
the density-of-state (DOS) \cite{tj1,tj7}. The energy position
of this gap depends on the length
of the supercell, i.e. on the wave length of the periodic potential
perturbation induced by the phonon/spin wave. Hence, if the one-particle
gaps appear at the Fermi energy, $E_F$, then there must be a correlation between
doping (determines $E_F$) and wave length. The appearance of the gap can be
understood from the nearly-free-electron model (NFE), which has also been
used to interpolate ab-initio results \cite{tj9}. The pseudogap is an important
part of cuprate physics, but the question if it helps or competes with
superconductivity is not settled. On the other hand, it can be argued that
artificial periodic waves such as given by periodic doping or oxygen ordering
will lead to higher $T_C$, if the waves are tuned correctly \cite{bianc,apl}.

Presented here is a many-particle extension of the NFE band model, which can be used for the simulation 
of the T-dependence of the pseudogap. 
The model is based on one-dimensional potential perturbations from
phonons or anti-ferromagnetic (AFM) spin waves. The
criterion for having gaps at $E_F$ is made from the estimated total energy (kinetic electronic energy and energy
of phonon/spin fluctuations) for an interacting electron-phonon (electron-spin) system,
with many-body coupling parameters $\lambda$ or $\lambda_{sf}$ for phonons or spinfluctuations,
respectively.
A continuous metallic band can be gapped
because of a periodic potential perturbation, as in the appearance of
a gap for semiconductors \cite{zim}. 
The potential perturbation is such that two regions of k-space, at $\vec{k}$
and -$\vec{k}$, are affected equally. Only
waves which modify electronic states around the Fermi surface (FS) are of interest,
since changes of the bands far from the Fermi energy ($E_F$) make no change in the
total kinetic energy. This puts a constraint on the $q$-vectors of the perturbations.
The results of the model imply similar mechanisms for the pseudogap and the superconducting gap
in high-$T_C$ cuprates.

%\section{Potential oscillations and band gaps.}

Phonons and spin fluctuations are normally excited thermally following the
Bose-Einstein occupation, $g(T,\omega)$, of the phonon- or spin wave density-of-states (DOS),
$F(\omega)$ or $F_m(\omega)$, respectively. 
The averaged atomic displacement amplitude for phonons, $u$, can be calculated 
as function of $T$ \cite{zim,grim}. Approximate results
make $u_Z^2 \rightarrow 3\hbar\omega_D/2K$ at low $T$
("zero-point motion", ZPM) 
and $u_T^2 \rightarrow 3 k_BT/K$ at high $T$
("thermal excitations"), where $\omega_D$ is a weighted average of $F(\omega)$. The force constant, $K=M_A\omega^2$,
where $M_A$ is an atomic mass, can be calculated as
$K = d^2E/du^2$ ($E$ is the total energy). The corresponding
relations for averaged fluctuation amplitudes of the magnetic moments, $m$,
are the same, but without the polarization factor 3 and with the replacement of $K$ 
with $K_m = d^2E/dm^2$ \cite{tjfe}. The force constants $K$ and $K_m$ do not change
with the oscillation amplitudes for harmonic oscillations.
The time scales of phonons
and electrons are sufficiently different for adiabatic relaxation of
electrons and band gaps.
The energy of an atomic oscillation, $U$, has an elastic contribution because of $u$, and a
kinetic contribution because of the velocity, $\upsilon$. The time dependence of the sum,
$2U(t) = K u^2 cos^2(\omega t) + M_A \upsilon^2 sin^2(\omega t)$,
is a constant in the harmonic approximation.  
$U(0)=\frac{1}{2}Ku^2$, where $u$ refers to the maximal atomic displacement, permits us to
calculate $U$ from static conditions, and to identify $Ku^2$ as
an ingredient of the standard expression for $\lambda$.
Acoustic phonons are more efficient for having a clear gap, 
while optical phonons sometimes have $u$=0 everywhere and a
smeared time average of the band gap.

Thermally activated
phonons and spin waves create some disorder and will change the electronic states 
and the DOS, $N(E)$ \cite{fesi}. Individual phonon- or spin-waves which
cause a gap close to $E_F$ in the normal state are particurarly interesting
for high-$T_C$ copper oxides \cite{tj1}.
The gap is largest in the CuO bond directions, while the FS remains sharp in the diagonal direction,
to form an FS-arc \cite{tj6}.
Band calculations show that atomic distortions of phonons 
create a periodic potential along a chain of atoms,
which can be modeled by
\begin{equation}
 V(x) = V_q e^{-i \vec{x}\cdot\vec{q}}
\label{eq1}
\end{equation}
for phonon propagation along $\vec{x}$ with wave vector $\vec{q}$ \cite{zim,tj1}.
The maximum amplitude, $V_q$, can be obtained from ab-initio band results as half of the
band gap, or as the maximum difference of the potential shift within the unit cell.
A spin wave makes an analogous perturbation
within the spin polarized part of the potential, where the densities of opposite spins
differ by a phase factor of $\pi$. The result is an 
AFM spin configuration with wave length given by $2\pi/q$.
Phonons and spin waves can be considered separately, but  
an important spin-phonon coupling (SPC) in the cuprates
leads to unusual properties \cite{tj6,tj7,egami}.
The following description, based on phonon excitations, can be adopted to spin waves and SPC.

The NFE band has a gap $E_g = 2 V_q$ at 
some "zone-boundary" $k=G/2$ in one dimension (1-D) \cite{zim,tj1}.
(All q-vectors are concerned initially, but a few of them are particular
because of energy gains from the gap at $E_F$.)
The general band dispersion as function of $k$

\begin{equation}
\varepsilon = \frac{1}{2}(k^2+(k-G)^2 \pm \sqrt{(k^2-(k-G)^2)^2+4V_q^2})
\label{eq3}
\end{equation}
can be made simpler near $E_F$ where $\epsilon \approx const. \cdot \kappa$, when the k-point, $\kappa$, 
is measured from the zone boundary.

 \begin{equation}
\varepsilon = \pm \sqrt{(\epsilon^2+V_q^2)}
\label{eq4}
\end{equation}
if $\epsilon$ is much smaller than the band width, $W$.
The wave function at $\kappa=0$ is commensurate with $V(x)$, while
for other k-points, when $\epsilon$ is below $E_F$, it is not.

%  The approximation of a linear $\epsilon$ as function of $\kappa$
% is valid for $\epsilon << W$, the band width from the bottom
% of the band to $E_F$. 
The normal free electron dispersion, $\varepsilon = \epsilon$,
is recovered for $V_q = 0$, and $N(\varepsilon) = N/|d\varepsilon/d\epsilon|$,
becomes constant and equal to $N$.
The gapped $\Tilde{N}$ is zero for $\epsilon=E_F \pm V_q$.
 At very low $T$ there is practically no thermal occupation of phonons. The Fermi-Dirac
occupation $f$ is essentially a
step function at $E_F$.

The propensity for having small or large gaps can be found through
ab-initio band calculations, since one-particle bands are more or less sensitive to the
potential perturbation. It depends on the material if this is going to be an important
effect or not. The total energy is reduced if a gap is formed at $E_F$ because of the
perturbation, but it is
not lower than the total energy for the system without phonon distortion. 
In the following it is argued that an additional effect, beyond the one-particle
band mechanism, can lead to spontaneous excitations of charge/spin wave gaps in systems
where $\lambda$ or $\lambda_{sf}$ are large. The $\lambda's$ are results of many-body
interactions across $E_F$, and they are not solely determined by the one-particle band structure.
The total energy for the vibrating system can then be lower than for a static system.

The electron-phonon coupling $\lambda$ is active for energies $\pm \hbar \omega$
around $E_F$, where it can be written $N M^2/K$ \cite{zim}.
The matrix
element  $M$ for energies inside the interval $\pm \hbar \omega$
 can be evaluated as $\langle \Psi^*(E_F,r) dV(r)/du \Psi(E_F,r) \rangle$,
which is the first order change in energy caused by the perturbation $dV(r)$ for $du \rightarrow 0$.
For a finite value of $u$ the change in energy will be finite and equal to the gap $V_q$, 
since $V_q/u$ is constant for harmonic vibrations. 
Thus, instead of calculating $M$ as a matrix element it is convenient to take the value directly
from the band gap, and $M$ can be written $V_q/u$ for energies close to $E_F$, which makes
$\lambda = N V_q^2/Ku^2$.

%\section{The band model at T=0.}
 
Totally there is a gain in energy if $|U| \leq |E|$. The system will
spontaneously increase $u$ of the vibrations (from the normal value given by thermal
excitations) in such a case. 
Other effects such as electron-electron
correlation and potential terms add to the energy costs 
and can prevent
a gap in many systems. 
With the DOS, $U$ and $E$ per unit cell
the condition $|U|=|E|$ is written

\begin{equation}
\frac{1}{2}Ku^2 = \int_{-\hbar\omega}^0 \epsilon (N(\epsilon)-\Tilde{N}(\epsilon)) d\epsilon 
\label{eqn4}
\end{equation}

for $T=0$, where $\Tilde{N}(\epsilon)$ is the DOS with the gap and ${N}(\epsilon)$, the DOS of the
normal state, is assumed
constant within $\hbar \omega$ around $E_F$. 
%(Vibrational amplitudes in the normal state will be included later).
The integration is to $\hbar \omega$ since 
$\lambda$ is zero for energies larger than $\pm \hbar \omega$.

 With a
substitution $e^2=\epsilon^2+V_q^2$ we obtain $\Tilde{N}=N|e|/\sqrt{e^2-V_q^2}$ and,

\begin{equation}
\frac{1}{2}Ku^2 = \int_{-\hbar\omega}^0 N \epsilon d\epsilon - \int_{-\hbar\omega}^{-V_q} Ne^2/\sqrt{(e^2-V_q^2)} de 
\label{eq5} 
\end{equation}

The result is
\begin{equation}
Ku^2 = N V_q^2 ln (2\hbar \omega/ V_q) 
\end{equation}

and

\begin{equation}
V_q = 2 \hbar \omega e^{-1/\lambda}
\label{eq6} 
\end{equation}

since $N V_q^2 / Ku^2$ turns out to be equal to $\lambda$.  
The coupling determines the gap through constant ratios of $V_q/u$.
This result is similar to the BCS equation for the superconducting gap \cite{bcs,bcstj}.
The present state is different from a normal gapped state obtained via standard band
calculations. It only appears because of electronic interaction near $E_F$, and it
disappears for small $\lambda$.
 
Eqn. \ref{eq5} has no solutions for $\hbar\omega < V_q$, so the
concurrent state with phonon softening, where all electronic energy goes into a renormalization of the phonon
with $K,\omega \rightarrow 0$, is not described. Neither using $W$ instead of $\hbar\omega$ in eqn \ref{eq5}
gives a proper description of a static case, because states far below $E_F$ do not have the correct wavelength
and cannot contribute to $V_q$.

%\section{The limit $V_q \rightarrow$0.}

The model for finding the maximal temperature for having a gap ($T^*$) is obtained from eq. \ref{eqn4}, but with
the Fermi-Dirac function, $f$,
as the $T$-dependent weight factor for $\Tilde{N}(\varepsilon)$
and $N(\varepsilon)$, and with the integration in the interval [$-\hbar\omega,\hbar\omega$].
At $T^*$ it is required that 
$V_q \rightarrow$ 0. This is solved numerically through integrations $I$;

\begin{eqnarray}
Ku^2 /N V_q^2 = 1/ \lambda = I(V_q) =\nonumber \\
(\int_{-\hbar\omega}^{\hbar\omega}  \epsilon f d\epsilon -
\int_{-\hbar\omega'}^{\hbar\omega'} e |e|/\sqrt{(e^2-V_q^2)} f de)/V_q^2 
\label{eqn10}
\end{eqnarray}

for $V_q \rightarrow$ 0. The ' in the second integral means that the energies where $|e| < V_q$ are excluded.

%The original BCS expression for $T_C$, which normally is derived from
%\begin{equation}
%1/gN = \int_0^{\hbar\omega} \frac{de}{e} tanh(e/2k_BT)
%\label{eqn11}
%\end{equation}
%(where $gN$ is the coupling constant \cite{fet}) is here solved numerically with the same
%precision as the expression above for very small $V_q$. 
The results, shown
in Fig. 1, has the same form as the solution for superconductivity:
% analytic solution of eq. \ref{eqn11}, the well-known BCS formula;

\begin{equation}
k_BT^* = 1.13 \hbar\omega e^{-1/\lambda}
\end{equation}

\begin{figure}
\includegraphics[height=6.0cm,width=8.0cm]{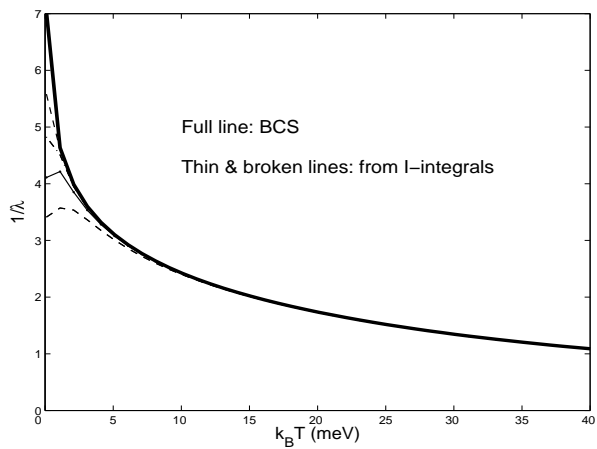}
\caption{The full line shows $1/\lambda$ as a function of $k_BT$
as obtained from the BCS result for $\hbar\omega$=0.1 eV. The thin and broken
lines show the corresponding result from eqn. \ref{eqn10}, for 5 different $V_q$'s
(0.5, 1, 2, 4 and 8 meV). The result for the smallest $V_q$ is indistinguishable
from the BCS result over this temperature range.
}
\label{fig1}
\end{figure}

 For example, it can be verified from this formula and Fig \ref{fig1} that a $\lambda$ of 0.5 makes
$k_BT^* \approx$ 15 meV when $\hbar\omega$ is 100 meV.

A non-constant DOS, with $N(\varepsilon)$ from eqn. \ref{eq1} inserted in eqn. \ref{eqn10},
and $\hbar\omega$ being a large fraction $W$, tends to decrease $T^*$.

For AFM spin waves there is a cost in magnetic energy, which in the harmonic approximation
can be written $U_m = \frac{1}{2}K_m m^2$. The change in potential on some site, $V_m$, is
positive for one spin and negative for the other spin direction. This defines a
$\lambda_{sf} = N V_m^2/K m^2$ as a coupling constant for spin-fluctuations \cite{tjfe}.
The rest of the equations are applicable with $\lambda_{sf}$ replacing $\lambda$ and
with $\hbar\omega_{sf}$ being the energy of the spin wave.

Typical atomic displacements and magnetic moments from phonons and spin waves in the
normal state can be determined from the effective force constants
$K$ and $K_m$ and the gap values,
$u=\sqrt(N/K\lambda)V_q$ and $m=\sqrt(N/K_m\lambda_{sf})V_q$, respectively.

%\section{Electric field.}

\begin{figure}
\includegraphics[height=6.0cm,width=8.0cm]{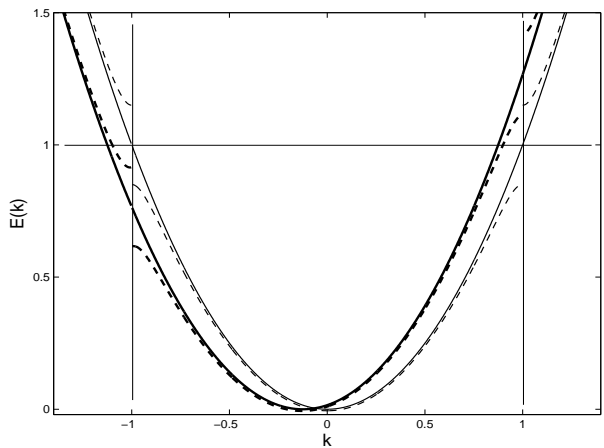}
\caption{Thin, full and broken lines; free electron band without and with a gap at
the zone limits (vertical lines), respectively. Bold full and broken lines: The bands modified by an 
electric field.  
}
\label{figq}
\end{figure}

The charge/spin wave state has low resistivity if the gap is complete over the FS, i.e. $\Tilde{N}(E_F)=0$.
An electric field $E_x$, applied during a time $\tau$, accelerates free electrons until their
velocity \.{x} becomes $eE_x \tau/m$, where $e$ and $m$ are the electron charge and mass, respectively.
The velocity $v_x(k)$ changes from $\hbar k /m$ to  $\hbar k /m + $\.{x} ($k$ is the momentum
along $\vec{x}$), and the free electron band from $\hbar^2 k^2/2m$ to $\hbar^2 k^2/2m +$ \.{x}k, as shown 
in fig. \ref{figq}. For a metal, with $E_F$ indicated by the horizontal line in fig. \ref{figq}, 
occupied states
near $k=+1$ ($k_F$) are pushed upward, while the states near $k=-1$ ($-k_F$) decrease their energy. The net
velocity $\int v_x(k) dk \approx 2 v_F$\.{x} and the conductivity is approximately $e^2 \tau N v_F^2/m$,
as from the Boltzmann equation \cite{zim,pba}.
But the total kinetic energy $E_t = \int N(\epsilon) \epsilon d\epsilon$ is enhanced by the field,
so the net current will decay through resistive dissipation when the field is switched off. 
A band insulator, with a gap at the
zone boundary, will also have a higher $E_t$
from an applied field, and by switching off the field, $E_t$ decays back to its low value for the ground state.

However, if
 the free electron band without field (thin line in fig. \ref{figq}) 
is shifted horizontally
by the electric field to a new position (bold line), 
then $E_t$ will remain at the minimum,
no decay is possible, so the state with the net current can persist. 
It is equivalent to moving the origin
of $k$ to $-2m/\hbar$\.{x}.
The electron occupation of the gapped band will 
not be asymmetric as the
bold broken line in fig. \ref{figq}, but  will be symmetric relative to the new origin as
 for the unshifted band. 
This can be achieved through phonons with slightly different momentum so that phonons
and electrons have a common drift.
(Two phonons with momentum $\pm P$ have energy $\hbar\Omega_P =
\sqrt{K/M_A} |sin(Pa)|$ and velocity $\pm a \sqrt{K/M_A} |cos(Pa)|$, where $a$ is the real space periodicity.
Together they have energy $2\hbar\Omega_P$ and zero velocity. If the phonons change
to $P+p$ and $-P+p$ (and $p << P$), there will be a net velocity, $V_p \approx 2a\sqrt{K/M_A}|sin(Pa)|pa$ =\.{x}.
The phonon energy, $2\hbar\Omega_P cos(pa)$, will not increase, since $cos(pa) < 1$.)

A common drift of electrons and phonons implies different lattice conduction  
along opposite directions relative to an imposed current, which also will be
different for a state with no current. Effects on lattice and spin-wave dependent 
properties should be observable although very small.

%\section{Magnetic field.}

From the selective q-dependence it can be shown that 
a weak magnetic field will destroy the pseudogap. As mentioned, one
$\vec{q}$ generates the gap on
the paramagnetic FS at $E_F$. A magnetic field, $H$, will split the FS into two, one for each
spin ("up" or "down"), and $E_F^{u,d} = E_F \pm \mu_BH$. Two independent phonons
are required for having the gaps optimally on two FS.
One has a potential perturbation at $e^{-i(\vec{q}-\vec{\delta})\cdot \vec{x}}$,
and the other at $e^{-i(\vec{q}+\vec{\delta})\cdot \vec{x}}$, where $\delta$ is
determined by the band dispersion and $\mu_BH$. The sum of these two potentials
is $2cos(\vec{\delta}\cdot \vec{x})e^{-i\vec{q}\cdot \vec{x}}$, and therefore,
even if there is a modulation given by the cosine function,
the effective $\vec{q}$ remains the same and cannot fit optimal values
for two FS.
The resulting gaps do not appear at $E_F$ on the two spin-split bands, which
will reduce the gain in energy.
The energy difference, $D(H,T)$, between the kinetic energy for the gapped DOS
with and without field, which is calculated as

\begin{equation}
D(H,T)=\int_{-\hbar\omega'}^{\hbar\omega'}e\Tilde{N}(e)(f(e+H,T)+f(e-H,T)-2f(e,T))de 
\label{eqnD}
\end{equation}
increases quadratically with the field amplitude
$H < V_q$ for low and high $T$ ($\approx V_q$). This  
is because the thermal occupation can be made more efficiently if $E_F$ is closer to the DOS peak (on $\Tilde{N}$) 
above the gap for "majority" and closer to the DOS peak below the gap in the "minority" states,
than if $E_F$ is in the middle of the gap. 

In conclusion, three mechanisms promote gaps at $E_F$. First, thermal normal state excitations
of phonons/spin waves generate potential perturbations and weak band gaps near $E_F$.
The total energy for the perturbed state is not lower than for the static state, but phonons
can be softened.
Second, systems with large $\lambda$ can spontaneously generate phonons/spin waves
because of a gain in total energy, as has been described above for NFE bands. The third mechanism is through
superconducting pairing. The equations for $T^*$ and $T_C$ are similar.
Also conductivity and the effect of a magnetic field behave in similar ways \cite{zim,kitt}. 
Calculations show increased
$\lambda$ for low doping, which would agree with the evolution of $T^*$. But the superconducting
$T_C$, which is maximal at a doping of 0.12-0.15 holes/Cu, does not follow the doping dependence of $\lambda$.
The shape of the FS and differences
in the $k,k'$-dependence of $\lambda$ (nesting etc.) are probably important for which state
will win.
In 3-D it is hard to imagine that a multitude of density waves can coexist to
make a full gap everywhere on the FS sphere. For 2-D cuprate bands it is simpler since $\lambda$ is
only large for k-points near $X$ and $Y$, where a pseudogap can appear. 
These results support the picture that
the pseudogap competes with superconductivity, and that it seems to depend on preformed pairs
between $k$ and $-k$ states.


\begin{thebibliography}{10}


%--------------
\bibitem{mull} J. Bednorz and  K.A. Muller, Z. Physik {\bf B64}, 189, (1986).

\bibitem{tall} J.L. Tallon and J.W. Loram, Physica C{\bf 349}, 53, (2001).

\bibitem{timu} T. Timusk and B. Statt, Rep. Prog. Phys. {\bf 62}, 61, (1999).

\bibitem{dama} A. Damascelli, Z.-X. Shen and Z. Hussain, Rev. Mod. Phys. {\bf 75}, 473, (2003).

\bibitem{tj1} T. Jarlborg, Phys. Rev. B{\bf 64}, 060507(R), (2001).

\bibitem{tj7} T. Jarlborg, Physica C{\bf 454}, 5, (2007).

\bibitem{tj9} T. Jarlborg, Phys. Rev. B{\bf 79}, 094530, (2009).

\bibitem{bianc} M. Fratini, N. Poccia, A. Ricci, G. Campi, M. Burghammer, G. Aeppli
and A. Bianconi, Nature {\bf 466}, 841, (2010).

\bibitem{apl} T. Jarlborg, Appl. Phys. Lett. {\bf 94}, 212503, (2009).

\bibitem{zim} J.M. Ziman, {\it Principles of the Theory of Solids} (Cambridge University
Press, New York, 1971).

\bibitem{grim} G. Grimvall, {\it Thermophysical properties of materials.}
(North-Holland, Amsterdam, 1986).

\bibitem{tjfe} T. Jarlborg, Physica C {\bf 385}, 513, (2003); Phys. Lett. {\bf A} 300, 518, (2002). 

\bibitem{fesi} T. Jarlborg, Phys. Rev. B{\bf 59}, 15002, (1999).

\bibitem{tj6} T. Jarlborg, Phys. Rev. B{\bf 76}, 140504(R), (2007).

\bibitem{egami} P. Piekarz and T. Egami, Phys. Rev. B{\bf 72}, 054530, (2005).

\bibitem{bcs} J. Bardeen, L.N. Cooper and J.R. Schrieffer,  Phys. Rev. {\bf 108}, 1175 (1957).

\bibitem{bcstj} An earlier version,  arXiv:0911.3079, (2009),
suggested that the superconducting state can be
simulated by the same model. This idea has been rejected by referees. 


\bibitem{pba} P.B. Allen, W.E. Pickett, and H. Krakauer, Phys. Rev. B{\bf 37}, 7482 (1988).

\bibitem{kitt} C. Kittel, "Introduction to Solid State Physics", 4th ed., Wiley, NY, (1971).

\end{thebibliography}
\end{document}